\documentclass[preprint,showpacs,superscriptaddress,preprintnumbers,amsmath,amssymb]{revtex4}

\usepackage{graphicx}
\usepackage{dcolumn}
\usepackage{bm}

\newcommand {\coseta}{\cos\theta_{\eta}^*}
\newcommand {\phieta}{\phi_{\eta}^*}
\newcommand {\etal} {{\it et al.}}

\newcommand {\msubx}{\ensuremath{M_{X}}}   

\begin{document}

\title{$Q^2$ dependence of the $S_{11}(1535)$ Photocoupling and Evidence for a $P$-wave resonance in $\eta$ electroproduction }

\newcommand*{\ABZ}{Abant Izzet Baysal University, Bolu 14280, Turkey}
\affiliation{\ABZ}
\newcommand*{\ANL}{Argonne National Laboratory, Argonne, Illinois 60439}
\affiliation{\ANL}
\newcommand*{\ASU}{Arizona State University, Tempe, Arizona 85287-1504}
\affiliation{\ASU}
\newcommand*{\UCLA}{University of California at Los Angeles, Los Angeles, California  90095-1547}
\affiliation{\UCLA}
\newcommand*{\CSU}{California State University, Dominguez Hills, Carson, CA 90747}
\affiliation{\CSU}
\newcommand*{\CMU}{Carnegie Mellon University, Pittsburgh, Pennsylvania 15213}
\affiliation{\CMU}
\newcommand*{\CUA}{Catholic University of America, Washington, D.C. 20064}
\affiliation{\CUA}
\newcommand*{\SACLAY}{CEA-Saclay, Service de Physique Nucl\'eaire, 91191 Gif-sur-Yvette, France}
\affiliation{\SACLAY}
\newcommand*{\CNU}{Christopher Newport University, Newport News, Virginia 23606}
\affiliation{\CNU}
\newcommand*{\UCONN}{University of Connecticut, Storrs, Connecticut 06269}
\affiliation{\UCONN}
\newcommand*{\DUKE}{Duke University, Durham, North Carolina 27708-0305}
\affiliation{\DUKE}
\newcommand*{\ECOSSEE}{Edinburgh University, Edinburgh EH9 3JZ, United Kingdom}
\affiliation{\ECOSSEE}
\newcommand*{\FU}{Fairfield University, Fairfield CT 06824}
\affiliation{\FU}
\newcommand*{\FIU}{Florida International University, Miami, Florida 33199}
\affiliation{\FIU}
\newcommand*{\FSU}{Florida State University, Tallahassee, Florida 32306}
\affiliation{\FSU}
\newcommand*{\GWU}{The George Washington University, Washington, DC 20052}
\affiliation{\GWU}
\newcommand*{\ECOSSEG}{University of Glasgow, Glasgow G12 8QQ, United Kingdom}
\affiliation{\ECOSSEG}
\newcommand*{\ISU}{Idaho State University, Pocatello, Idaho 83209}
\affiliation{\ISU}
\newcommand*{\INFNFR}{INFN, Laboratori Nazionali di Frascati, 00044 Frascati, Italy}
\affiliation{\INFNFR}
\newcommand*{\INFNGE}{INFN, Sezione di Genova, 16146 Genova, Italy}
\affiliation{\INFNGE}
\newcommand*{\ORSAY}{Institut de Physique Nucleaire ORSAY, Orsay, France}
\affiliation{\ORSAY}
\newcommand*{\ITEP}{Institute of Theoretical and Experimental Physics, Moscow, 117259, Russia}
\affiliation{\ITEP}
\newcommand*{\JMU}{James Madison University, Harrisonburg, Virginia 22807}
\affiliation{\JMU}
\newcommand*{\KYUNGPOOK}{Kyungpook National University, Daegu 702-701, South Korea}
\affiliation{\KYUNGPOOK}
\newcommand*{\MIT}{Massachusetts Institute of Technology, Cambridge, Massachusetts  02139-4307}
\affiliation{\MIT}
\newcommand*{\UMASS}{University of Massachusetts, Amherst, Massachusetts  01003}
\affiliation{\UMASS}
\newcommand*{\MOSCOW}{Moscow State University, Skobeltsyn Nuclear Physics Institute, 119899 Moscow, Russia}
\affiliation{\MOSCOW}
\newcommand*{\UNH}{University of New Hampshire, Durham, New Hampshire 03824-3568}
\affiliation{\UNH}
\newcommand*{\NSU}{Norfolk State University, Norfolk, Virginia 23504}
\affiliation{\NSU}
\newcommand*{\OHIOU}{Ohio University, Athens, Ohio  45701}
\affiliation{\OHIOU}
\newcommand*{\ODU}{Old Dominion University, Norfolk, Virginia 23529}
\affiliation{\ODU}
\newcommand*{\PITT}{University of Pittsburgh, Pittsburgh, Pennsylvania 15260}
\affiliation{\PITT}
\newcommand*{\RPI}{Rensselaer Polytechnic Institute, Troy, New York 12180-3590}
\affiliation{\RPI}
\newcommand*{\RICE}{Rice University, Houston, Texas 77005-1892}
\affiliation{\RICE}
\newcommand*{\URICH}{University of Richmond, Richmond, Virginia 23173}
\affiliation{\URICH}
\newcommand*{\SCAROLINA}{University of South Carolina, Columbia, South Carolina 29208}
\affiliation{\SCAROLINA}
\newcommand*{\JLAB}{Thomas Jefferson National Accelerator Facility, Newport News, Virginia 23606}
\affiliation{\JLAB}
\newcommand*{\UNIONC}{Union College, Schenectady, NY 12308}
\affiliation{\UNIONC}
\newcommand*{\VT}{Virginia Polytechnic Institute and State University, Blacksburg, Virginia   24061-0435}
\affiliation{\VT}
\newcommand*{\VIRGINIA}{University of Virginia, Charlottesville, Virginia 22901}
\affiliation{\VIRGINIA}
\newcommand*{\WM}{College of William and Mary, Williamsburg, Virginia 23187-8795}
\affiliation{\WM}
\newcommand*{\YEREVAN}{Yerevan Physics Institute, 375036 Yerevan, Armenia}
\affiliation{\YEREVAN}
\newcommand*{\NOWOHIOU}{Ohio University, Athens, Ohio  45701}
\newcommand*{\NOWINDSTRA}{Systems Planning and Analysis, Alexandria, Virginia 22311}
\newcommand*{\NOWUNH}{University of New Hampshire, Durham, New Hampshire 03824-3568}
\newcommand*{\NOWCUA}{Catholic University of America, Washington, D.C. 20064}
\newcommand*{\NOWSANPAULO}{San Paulo University, Brazil}
\newcommand*{\NOWUMASS}{University of Massachusetts, Amherst, Massachusetts  01003}
\newcommand*{\NOWMIT}{Massachusetts Institute of Technology, Cambridge, Massachusetts  02139-4307}
\newcommand*{\NOWECOSSEE}{Edinburgh University, Edinburgh EH9 3JZ, United Kingdom}
\newcommand*{\NOWGEISSEN}{Physikalisches Institut der Universitaet Giessen, 35392 Giessen, Germany}

\author {H.~Denizli} 
\affiliation{\PITT}
\affiliation{\ABZ}
\author {J.~Mueller} 
\affiliation{\PITT}
\author {S.~Dytman} 
\affiliation{\PITT}
\author {M.L.~Leber} 
\affiliation{\PITT}
\author {R.D.~Levine} 
\affiliation{\PITT}
\author {J.~Miles} 
\affiliation{\PITT}
\author {K.Y.~Kim} 
\affiliation{\PITT}
\author {G.~Adams} 
\affiliation{\RPI}
\author {M.J.~Amaryan} 
\affiliation{\ODU}
\author {P.~Ambrozewicz} 
\affiliation{\FIU}
% ...
\collaboration{The CLAS Collaboration}
     \noaffiliation
%
 
%
%

 


\date{\today}

\begin{abstract}
New cross sections for the reaction $ep \rightarrow e'\eta p$ are reported
for total center of mass energy $W$=1.5--2.3 GeV and invariant squared 
momentum transfer $Q^2$=0.13--3.3 GeV$^2$.  This large kinematic range allows
extraction of new information about response functions, 
photocouplings,
and $\eta N$ coupling strengths of baryon resonances.  A sharp structure 
is seen at $W\sim$ 1.7 GeV.  The shape of the differential cross section 
is indicative of the presence of a $P$-wave resonance that persists 
to high $Q^2$.
Improved values are derived for the photon coupling amplitude for the
$S_{11}$(1535) resonance.  The new data greatly expands the $Q^2$ range
covered and an interpretation of all data with a consistent
parameterization is provided.
\end{abstract}

                                                                                
\maketitle

\section{\label{sec:intro}Introduction}
Photoproduction and electroproduction experiments on the nucleon 
provide a clean probe
 of nucleon structure since Quantum Electrodynamics is well understood.
As a result, the matrix elements for $\gamma N \rightarrow N^*,\Delta^*$ transitions,
 commonly called the photon coupling amplitudes, are sensitive to 
 the nucleon and $N^*$ quark-level wave function.
These amplitudes have traditionally been calculated using quark 
models~\cite{Konen:1989jp,Capstick:1995ne,Close:1990aj}, 
but recently progress has been made in applying the techniques of Lattice 
QCD~\cite{Alexandrou:2004xn,Zanotti:2003gc,Basak:2006fq}.
Experimental measurements are currently being made of a number of 
different baryon resonances in several different final states.
For a review of the current status, see Reference ~\cite{Krusche:2003ik,Burkert:2004sk}.

Disentangling the wide and overlapping states that populate reaction data has
been a long-lasting problem.  In the mass region above total center of
mass (c.m.) energy ($W$) of 1.5 GeV, many overlapping baryon states are present
 and some are not well known.
The reaction $e p \to e' \eta p$ is especially clean, since processes
involving $\eta N$ final states
couple only to isospin $\frac{1}{2}$ resonances, simplifying the analysis.
A prominent peak in the total cross section is
seen for $\eta$ production at $W=1.535$ GeV in both $\gamma N$ and $\pi N$ experiments.
This is widely interpreted as the excitation of
a {\it single} resonance, the spin $\frac{1}{2}$, negative parity,
isospin $\frac{1}{2}$ state $S_{11}$(1535)~\cite{PDG:2006}.  
This state has a branching ratio to $\eta N$ of 45-60\% compared to at most
a few percent~\cite{PDG:2006,Vrana:1999nt} for other states.  This is a very
interesting and unusual pattern.  

Eta photoproduction experiments have
reaffirmed the strong energy dependence and $S$-wave (isotropic) character
close to threshold~\cite{Krusche:1995nv}.  Using polarized
photons~\cite{Ajaka:1998zi}, new values for $\eta N$ decay branching
ratios of other resonances have been determined through interference 
with the dominant $S_{11}$(1535).

Electroproduction cross sections can be used to  
extract the photocoupling amplitude for non-zero values of the squared momentum 
transfer ($Q^2$) from the electron to the resonance.
Using $\eta$ electroproduction~\cite{Kummer:1973em, Beck:1974wd, Alder:1975tv, Brasse:1978as, Breuker:1978qr, Brasse:1984vm, Armstrong:1998wg}, 
an unusually flat $Q^2$ dependence of the photocoupling amplitude was found for
the $S_{11}$(1535) in contrast to the nucleon form factors and photon
coupling amplitudes of other established resonances,
e.g.\ $P_{33}$(1232)~\cite{Ungaro:2006df}.  Although previous $\eta$ angular distributions
were largely isotropic at all $Q^2$, no detailed
response functions were extracted because of the poor
angular coverage using traditional magnetic spectrometers.
Although the $Q^2$ dependence was clearly different than for other
resonances, the results were comprised of many different 
experiments whose results appeared to be inconsistent with each other.
An analysis by Armstrong \etal~\cite{Armstrong:1998wg} showed that 
much of the inconsistency was due to different assumptions about
$S_{11}$(1535) properties used by the individual experiments.

In our previous publication~\cite{Thompson:2000by} we presented results 
on $\eta$ electroproduction
based on the first data taken with the 
CEBAF Large Acceptance Spectrometer 
(CLAS)~\cite{Mecking:2003zu} at Jefferson Lab.  We extracted the 
photocoupling amplitude ($A_{\frac{1}{2}}$) for the $S_{11}(1535)$ over the 
range 0.25 GeV$^2<Q^2<$1.5 GeV$^2$ from our data.
In addition, we observed indication of a structure at $W\approx 1.7$ GeV 
in the total cross section which is also seen as a change in the shape of 
the differential cross section at the same energy.  

The energy region around $W\approx 1.7$ GeV has received significant
attention lately.  A CLAS $\pi^+ \pi^-$ electroproduction 
experiment~\cite{Ripani:2002ss} 
found excess strength at the same energy beyond theoretical predictions 
based on previous data.
This excess strength was tentatively identified as a $P$-wave resonance;
either the decay properties of $P_{13}$(1720) change significantly 
or there is a new spin $\frac{3}{2}^+$ state. 
A recent $\eta$ photoproduction experiment at Bonn~\cite{Crede05} provides
a comprehensive set of cross section data from near threshold to
well beyond the resonance region.  
In the same paper~\cite{Crede05} a partial wave analysis of these
and other $\eta$ electroproduction data finds strong excitation of 
a $J=3/2^+$ state at 1775 MeV which they identify with the $P_{13}$(1720).  
Although previous analyses~\cite{PDG:2006} found weak evidence for any 
$\eta N$ decay of resonances at $W\sim$1.7 GeV, the new data is of
much higher quality than the older data.

The data and analysis reported here use a data set taken with
the same apparatus (CLAS) as used in our first 
publication~\cite{Thompson:2000by}.  The new
data have an order of magnitude more $\eta$ events than in that previous
paper  and a much larger kinematic range.  Therefore, the
new values presented here supersede the previously published 
data.  Our reach in $Q^2$ (0.13 {\rm GeV}$^2$--3.3 {\rm GeV}$^2$) is more than twice as 
large as in our first publication.  This allows a large extension of 
the $Q^2$ range where we can extract the photocoupling amplitude 
$A_{\frac{1}{2}}$ of the proton to $S_{11}(1535)$ transition.
We also more precisely determine the non-isotropies in the 
differential cross section and show evidence for a significant 
contribution to $\eta$ electroproduction due to a $P$-wave resonance
with a mass around 1.7 GeV.

The paper first presents some formalism needed to understand
the measurement and its analysis, followed by details of the experiment.
We then present results for the inclusive and exclusive analyses.
Discussion of these results with a Breit-Wigner model and conclusions
complete the paper.

\section{\label{sec:form}Formalism}
\begin{figure}[ht]
  \includegraphics[width=8.5cm]{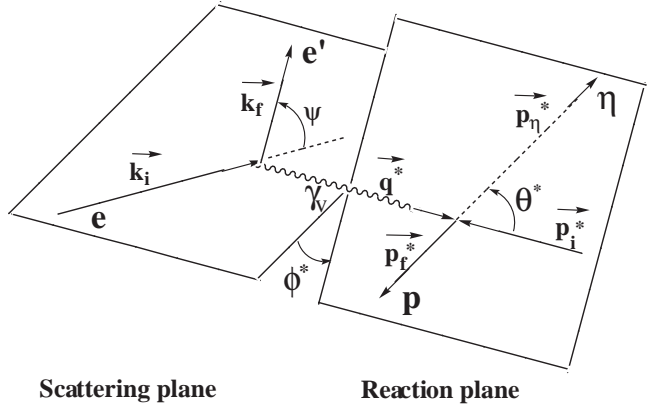}
  \caption{The reaction is depicted in the center of mass system, where the resonance is at rest.  The meson decay polar angle is defined relative to the virtual photon momentum, and the azimuthal angle is defined relative to the the electron scattering plane.}
  \label{fig:eptoeta}
\end{figure}

The kinematics for the $ep\to e' \eta p$ reaction are shown 
in Figure \ref{fig:eptoeta}.  It can be characterized in terms of the 
squared 4-momentum transfer between the electron and proton ($-Q^2$) 
carried by the virtual photon ($\gamma_v$), the invariant mass of 
the $\gamma_v-p$ system ($W$), and the scattering angles of the 
final state $\eta$ in the rest frame of the $\gamma_v-p$ system 
($\theta^*$, $\phi^*$).  These angles are also the decay angles
of the resonance in its rest frame.  We use the superscript * for 
quantities evaluated in this frame.
The five-fold unpolarized differential cross section for the 
$ep\to e' \eta p$ process at a specific energy ($E$) may be expressed 
as the product of the transverse virtual photon flux 
($\Gamma_\gamma$) in the Hand convention~\cite{Hand:1963bb} and the c.m. 
cross section for virtual-photoproduction of the $p\eta$ pair:
\begin{equation}
        \frac{d^{5}\sigma}{dWdQ^2d\Omega_{\eta}^*} = \Gamma_{\gamma}(E,W,Q^2)\frac{d^{2}\sigma}{d\Omega_{\eta}^{*}} (\gamma_v p \longrightarrow \eta p ).
\end{equation} \label{eqn:xs}
The cross section for the virtual reaction $\gamma_v p \longrightarrow \eta p$ is written by convention to explicitly display the dependence on $\phi^*$.
\begin{eqnarray}\label{eqn:equ1}
    \frac{d^{2}\sigma}{d\Omega_{\eta}^{*}}(\gamma_v p \longrightarrow \eta p )
   &=&\sigma_T + \epsilon\sigma_L +\nonumber\\ 
   & & \sqrt{2\epsilon(1+\epsilon)} \sigma_{LT}{\cos{\phieta}} +\nonumber\\
   & & \epsilon \sigma_{TT}{\cos{2\phieta}}, 
\end{eqnarray}
where $\epsilon$ is the longitudinal degree of polarization of the virtual photon and is given by
\begin{equation}
  \epsilon = \left[1+2(1+\frac{q^2}{Q^2})\tan^2 \left(\frac{\Psi}{2} \right)\right]^{-1},
\end{equation}
where $q$ is the magnitude of the 3-momentum of the virtual photon and $\Psi$ is the electron scattering angle.
Since $\epsilon$ is invariant under collinear transformations, $q$ and 
$\Psi$ may be expressed either in the lab or c.m. frame.  The cross sections 
for transverse and longitudinal photon are represented by
$\sigma_T$ and $\sigma_L$, respectively.
In addition, $\sigma_{LT}$ is a contribution due to interference between 
transverse and longitudinal amplitudes and $\sigma_{TT}$ describes the 
interference between amplitudes for the two different transverse 
polarizations, either aligned or anti-aligned with the spin of the target 
proton. All four of these terms depend on $W$, $Q^2$, and $\cos\theta^*$.

In order to identify individual baryon resonances, the cross section 
should be decomposed into partial wave amplitudes. 
These amplitudes are most often labeled by the electromagnetic 
multipole notation~\cite{CGLN}.
Multipoles are commonly labeled $E_{l\pm}$, $M_{l\pm}$, and $S_{l\pm}$,
where $l$ is the orbital angular momentum of the final $\eta p$ system and
$\pm$ denotes whether the total angular momentum is $l\pm\frac{1}{2}$.
$E$ and $M$ refer to electric and magnetic transitions involving 
transverse virtual photons, while the longitudinal ($S$) transitions involve 
longitudinal photons.

The response functions and multipoles have contributions from 
underlying resonant and nonresonant reaction mechanisms. 
When evaluated at the peak of the resonance, the multipole 
is expressed in terms of both the photocoupling 
amplitude and the hadronic decay properties of the individual 
resonances in a commonly accepted way~\cite{PDG76}.
The photocoupling amplitudes are
labeled by the $\gamma N$ total helicity ($\frac{1}{2}$ or $\frac{3}{2}$) 
and the virtual photon
polarization (transverse or longitudinal) and depend on the
invariant squared momentum transfer to the resonance ($Q^2$).  
The shape of the resonance determines the $W$ dependence of
the resonant part of the multipole.
Spin-$\frac{1}{2}$ resonances will be described by one transverse
amplitude ($A_{\frac{1}{2}}$) and one
longitudinal amplitude ($S_{\frac{1}{2}}$).
In terms of multipoles, an $S_{11}$ resonance has an 
$E_{0+}$ (electric dipole) and a $S_{0+}$ transition; 
a $P_{11}$ has $M_{1-}$ and $S_{1-}$ transitions.
Higher spin resonances will be described by both 
$A_{\frac{1}{2}}$ and $A_{\frac{3}{2}}$
photocouplings (and both $E$ and $M$ multipoles).  Extraction
of the multipole amplitudes from the cross section data, see
e.g. \cite{Workman02}, is not unique because more than one
bilinear combination of multipoles have identical angular
distributions.  We therefore choose simplified
methods (discussed below) to analyze the data.

The differential cross sections can be calculated with a model
of resonance production/decay and the nonresonant processes.
This cannot yet be done from a fundamental field theory such
as Quantum Chromodynamics (QCD).  Instead,
models are used that have parameters determined from
data. The $\eta$-MAID~\cite{Chiang:2001as} model uses an isobar 
model~\cite{Drechsel:1998hk} to
construct the cross section for $\eta$ photo- and electroproduction;
parameters are fit by comparisons with
previous results~\cite{Krusche:1995nv,Thompson:2000by,Dugger:2002ft}.
We have calculations from the MAID code for our kinematics.  To further
understand our data, we also do Legendre polynomial fits
to the angular distributions.
Both these results are described in Sect.~\ref{sec:differ}.

To analyze our angle-integrated cross sections, we make a further 
simplification which is possible because the $S_{11}$(1535) resonance
is dominant near threshold.  Therefore, we ignore the nonresonant
amplitude.
If one can isolate the contribution of a single $S_{11}$ resonance 
to the $E_{0+}$ multipole, the cross section takes a simple form,
\begin{equation}
\frac{d\sigma}{d\Omega_{\eta}^*}= \frac{p_\eta^* W}{m_p K}|E_{0+}(W)|^2,
\label{eq:xs}
\end{equation}
where $K=(W^2-m_p^2)/(2m_p)$ is the equivalent real photon energy,
$p^*_\eta$ is the momentum of the outgoing 
$\eta$ in the $S_{11}$ rest frame, and $m_p$ is the proton mass.  The 
longitudinal multipole, $S_{0+}$, in principal contributes and we
do not have the data to make the separation.  However, $S_{0+}$ has
been found to be small~\cite{Breuker:1978qr},
and it was therefore ignored in previous analyses~\cite{Kummer:1973em,Beck:1974wd,Brasse:1978as,Breuker:1978qr,Brasse:1984vm,Krusche:1995nv,Armstrong:1998wg,Thompson:2000by}.  An isobar model analysis of $\eta p$, $\pi^0 p$, and
$\pi^+ n$ CLAS electroproduction data~\cite{Aznauryan:2005tp} confirms the
assumption of a small longitudinal component.  In this analysis, the
value of $S_\frac{1}{2}/A_\frac{1}{2}$ is about 15-20\%; this
translates to a few percent contribution to the cross sections measured here. 
With the assumption that a single resonance dominates the cross
section and $S_\frac{1}{2}$ is small, $A_{\frac{1}{2}}$ for 
$\gamma p \rightarrow S_{11}$(1535) can be determined from~\cite{PDG76}
\begin{equation}
  A_{\frac{1}{2}}=\sqrt{2\pi \frac{p^*_\eta W_R^2 \Gamma_R}{Km_p^2 b_\eta}}\mathrm{Im}(E_{0+}(W_R)),
\label{eq:a12}
\end{equation}
where $E_{0+}$ refers only to the contribution from the resonance which
is evaluated at the peak of the resonance.
If there are other contributions to $E_{0+}$, a model is needed to extract the resonance contribution.
This formula contains terms related to the final state decay of the $S_{11}$: 
$\Gamma_R$ (the total width of the $S_{11}(1535)$), $b_\eta$ (the branching 
fraction into the $\eta p $ final state), and $p^*_\eta$ are all calculated 
at the mass of the $S_{11}$ ($W_R$).
Our current lack of knowledge of these parameters leads to a 
model dependence in the extracted values for $A_{\frac{1}{2}}$.
This prompted Benmerrouche \etal~\cite{Benmerrouche:1995uc} to propose 
using a quantity for each resonance with
less model dependence
\begin{equation}\label{equ:xi}
\xi_\frac{1}{2}=\sqrt{\frac{m_p^2 K b_\eta}{W^2_R p^*_\eta\Gamma_R}}A_\frac{1}{2}.
\end{equation}
$A_\frac{1}{2}$ depends on the matrix element for the initial state $\gamma N \rightarrow N^*$ transition, while $\xi_\frac{1}{2}$ is proportional to the product of the matrix elements for the $\gamma N \rightarrow N^*$ and  $N^* \rightarrow\eta p$ transition.  For $S_{11}$(1535), $\xi_\frac{1}{2}=\sqrt{2\pi}\mathrm{Im}(E_{0+}(W_R))$.
Although $\xi_\frac{1}{2}$ is more closely related to experimental values, $A_\frac{1}{2}$ is more easily determined from calculations, e.g. using quark models. 
Whichever quantity is used, the model dependence still exists when comparing
calculations to experiment.

We use the same resonance parameterization in all our calculations.
The relativistic Breit-Wigner form is taken from previous $\eta$ 
photoproduction work~\cite{Krusche97,Knochlein:1995qz} and extended to 
non-zero angular momentum as:
\begin{equation}
\frac{d\sigma_{BW}}{d\Omega_{\eta}^*}(W) = \frac{p_\eta^*}{q^*} \left|T_{BW}^\ell(W) \right|^2
\label{eq:sigbw}
\end{equation}
\begin{equation}
T_{BW}^\ell(W) = \frac{a W_R \Gamma_\eta}{(W_R^2 - W^2)-iW_R\Gamma_{tot}},
\label{eq:bw}
\end{equation}
where $a$ is a constant that contains the photon coupling amplitude
and kinematic factors, $q^*$ is the photon 3-momentum 
in the resonance rest frame, $\Gamma_\eta$ is the partial width for 
$N^*\rightarrow \eta p$ decay, and $\Gamma_{tot}$ is the total width,
\begin{equation}
\Gamma_\eta = \frac{B_\ell(p_\eta^*)}{B_\ell(p_{\eta,R}^*)}\Gamma_R
\end{equation}
\begin{equation}
\label{eq:width}
\Gamma_{tot} = \left(0.5\frac{p_\eta^*}{p_{\eta,R}^*}
\frac{B_\ell(p_\eta^*)}{B_\ell(p_{\eta,R}^*)} + 
0.4\frac{p_\pi^*}{p_{\pi,R}^*}
\frac{B_\ell(p_{\pi}^*)}{B_\ell(p_{\pi,R}^*)} + 0.1 \right) \Gamma_R.
\end{equation}
$\Gamma_R$ is the bare width and  $B_\ell(p^*)$ is a Blatt-Weisskopf 
penetration factor~\cite{BW}.  If $\ell$=0, this factor is equal to 1.
The momentum ratios~\cite{Knochlein:1995qz} approximately account 
for proper phase space effects for the various final states 
($\pi N$, $\eta N$, and $\pi \pi N$ where the phase space factors
for the $\pi \pi N$ final state are ignored).  They are weighted according
to estimates of the branching fractions to each final state.
This form has been successful in matching data, but is not unique.

\section{\label{sec:anal}Detector and Analysis}
The CLAS facility~\cite{Mecking:2003zu} was designed for efficient detection
of multi-particle final states.  The data used for this measurement
was taken in 1999 at electron beam energies of 1.5, 2.5 and 4.0 GeV.  
A cylinder of liquid hydrogen was used as the target.  Two different 
targets were used, 5.0 and 3.8 cm long.
Toroidal magnet coils separate CLAS into 6 largely identical sectors,
 each covering roughly $54^\circ$ in azimuthal angle $\phi$ 
(with smaller coverage at smaller polar angle).  Tracking drift chambers 
(DC) in CLAS measure angles and momenta of charged particles
 for lab polar angles in the range $8^\circ<\theta<142^\circ$.
Outside the DC, scintillation counters (SC) provide time-of-flight measurements
 with which we can separate the charged hadrons into pions, kaons, and protons.
For lab angles $\theta<48^\circ$, threshold \v{C}erenkov counters (CC) and
 Electromagnetic Calorimeters (EC) distinguish electrons from charged hadrons.

For this analysis, events were selected with an identified electron and
  proton.
Since the momentum 4-vectors of the beam and target are known, the 4-vector for the putative $\eta$ can be determined from these two final state particles.
A fiducial cut on these particles was applied to avoid the regions
 near the magnetic coils and the edges of the CC where the acceptance
is changing rapidly.  
The momentum of the electron was required to be above 400 MeV in order to be
 well above the trigger threshold.

Cross sections were calculated as a function of $Q^2$ and $W$ for
the angle-integrated data analysis and as a function of $Q^2$, $W$, $\cos\theta^*$,
and $\phi^*$ for the differential data analysis.  Cross sections
are determined in a standard way by determining the yield in each
of many bins, correcting for detector acceptance, and normalizing
by the beam intensity measured with a Faraday cup and the
calculated target thickness.

As discussed in Sect.~\ref{sec:form}, the extraction of resonance
properties comes from an analysis of the $\cos\theta^*$, $\phi^*$
distributions at specific values of $W$ and $Q^2$.  Distributions
of these variables covered by the apparatus are determined by
geometry.  The large acceptance of CLAS guarantees almost complete 
coverage in $\cos\theta^*$ and $\phi^*$.  The beam energies of the
experiment coupled with CLAS provided data at a wide range of $Q^2$
and $W$.  We compare the kinematic range for the new experiment
with that available for the previously published $\eta$ electroproduction
data in Table~\ref{tab:prev_kine}.

\begin{table}[!ht]
\begin{center}
\begin{tabular}{|l|c|c|c|c|} \hline
Experiments & W (GeV)       & Q$^2$ (GeV$^2$)   & cos$\theta^*$     &$\phi^*$ (deg) \\ \hline
Daresbury\cite{Kummer:1973em}    & $1.51\to1.55$   & $0.15\to1.5$      &   not given            & not given  \\
Bonn\cite{Beck:1974wd} & $1.51\to1.56$ & $0.2 \to 0.4$  & $-0.766\to0.939$& $\sim$0     \\
DESY\cite{Alder:1975tv}     & $1.5\to1.7$   & $0.22 \to 1.0$  & $-1\to1$  & $0\to180$ \\
DESY\cite{Brasse:1978as} & $1.49\to1.58$ & $0.6, 1.0$   & $-1\to1$        & $15\to90$ \\
Bonn\cite{Breuker:1978qr}  & $1.44\to1.64$ & $0.4$      & $-0.643\to0.866$& $-40\to40$    \\
DESY\cite{Brasse:1984vm}  & $1.49\to1.8$  & $2.0, 3.0$  & $-1\to1$        & $0\to120$  \\
JLab\cite{Armstrong:1998wg}  & $1.48\to1.62$ & $2.4, 3.6$ & $-1\to1$        & $0\to360$ \\
JLab\cite{Thompson:2000by}  & $1.5\to1.86$  & $0.375 \to 1.5$ & $-1\to1$    & $0\to360$ \\
this experiment            & $1.5\to2.3$  & $0.13 \to 3.3$ & $-1\to1$ & $0\to360$ \\
\hline
\end{tabular}
\caption{Summary of the kinematic ranges of previously published data
compared to this experiment.  This experiment
is an extension of Ref.~\cite{Thompson:2000by}.
Values given are the maximum ranges for each experiment.}
\label{tab:prev_kine}
\end{center}
\end{table}

\begin{figure}[ht]
  \includegraphics[width=8.5cm]{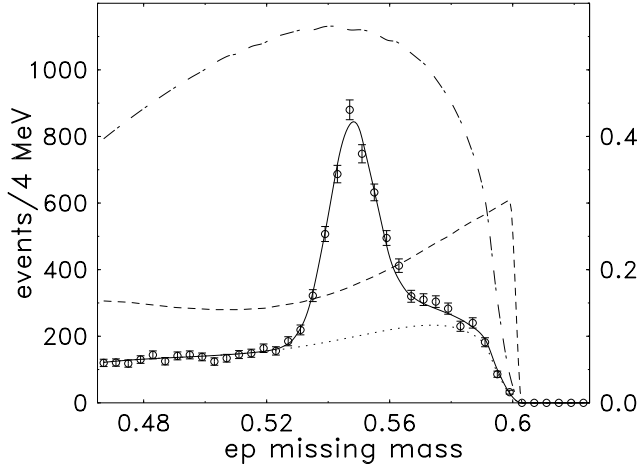}
  \caption{Sample missing mass ($M_X$) spectrum for $e p \rightarrow e p X$.  
The bin shown is for $W$=1.535 GeV and $Q^2$= 0.6 GeV$^2$. The dashed line 
(right scale) shows the acceptance that is calculated with a Monte
Carlo program. The sum of the $\eta$ signal shape and the raw background 
function modified by the acceptance function is then fit to the missing 
mass spectrum for each bin.  In the figure, the dot-dashed curve is the 
raw background function, $D_{bkg}$ from this fit, while the dotted curve 
shows those same values when multiplied by the acceptance function.  The solid 
curve shows the full fit.}
  \label{fig:mmfit}
\end{figure}
Events were divided into separate kinematic bins as detailed in
 Sec. \ref{sec:rslt}.
For each bin the $\eta$ yield was determined by fitting the distribution 
of missing mass recoiling against the outgoing $e$-$p$ system.
An example fit in one bin is shown in Figure \ref{fig:mmfit}.

The fit is the sum of a signal at the $\eta$ mass and a background
function.  The signal shape has a radiative tail and is corrected
for experimental resolution; the background function is a polynomial.
We use the data to determine both shapes.
Both functions must then be modified by the geometric acceptance for this
 reaction because it has a rapid variation with respect to the
kinematic parameters.  This method is an extension of what was used in the
previous CLAS data analysis~\cite{Thompson:2000by}.  

The shape of the signal was modeled in 2 steps to reproduce all
features seen in the data. 
It is first described by a delta function at the $\eta$ mass ($m_\eta$) plus an exponential above  $m_\eta$ representing the radiative tail,
$$S(m) = (1-f) \delta(m-m_\eta) + f \Theta(m-m_\eta) e^{-\alpha(m-m_\eta)}.$$
The fraction of events in the radiative tail ($f$), and a parameter
describing the slope in 
the exponential ($\alpha$) were determined for each $W$-$Q^2$ bin from 
Monte Carlo generator events containing radiative 
effects~\cite{Ent01,MoTsai69}.  
This signal shape was then convoluted 
with a Gaussian representing the experimental resolution to obtain the 
final signal shape (an analytic function) used to fit for the $\eta$ yield.
In the fit to obtain the yields, only the magnitude of the signal
and background functions were free.  All other parameters were determined
separately.
The rms resolution for the missing mass peak ranged from 4 MeV in the 
low $Q^2$ bins up to 12 MeV at the highest $Q^2$.  The experimental $\eta$ 
mass, experimental resolution width, $f$, and $\alpha$ were first fit
to simple functions of $Q^2$ and $W$ for
both data and Monte Carlo to smooth out statistical variations.
The experimental $\eta$ mass was found to be within 1 MeV of the
accepted value.  Estimated contributions to the systematic uncertainty by 
these choices of parameters were evaluated in a later step.

The background comes from $\pi \pi$ production.  Although it has a smooth
dependence on $M_X$, no models are available.
We fit the background with a simple polynomial
($D_{bkg}$) in the missing mass ($M_X$) constrained to be zero at the 
kinematic limit ($m_{max}$) as required by the decreasing phase-space,

\begin{equation}
        D_{bkg}(M_X) = b_{0}\left( 2\sqrt{\Delta m' \Delta m} - \Delta m \right),
\end{equation}
where $\Delta m$ $=$ $m_{max} - M_X$ and $b_0$ is the overall strength.
One example of this function is shown in Fig.~\ref{fig:mmfit}.
At the highest beam energy, a slightly more complicated function was used.
Both forms contained one parameter ($\Delta m'$) that was determined from
our data by fitting to a polynomial in $W$.  As with the peak shape function
parameters, these fit parameters were included in the systematic uncertainty 
determination.

The main structure in the background fit function 
comes from the variation of the geometric acceptance of the detector as 
a function of missing mass.  We found that proper modeling of this acceptance
was very important.  We determined this acceptance using a separate
Monte Carlo program that generated 
$e p \to e p X$ events with the $X$ mass thrown randomly across the 
fit region.  After requiring the scattered electron and proton to be in 
the fiducial volume of CLAS, we compared the generated and accepted events 
in order to calculate the background acceptance function.
We multiply our simple background function with the calculated acceptance 
function to obtain the final background used in the fit.
Examples of all three of these curves are shown in Fig.~\ref{fig:mmfit}.

In a small number of bins where the cross section is low, statistical fluctuations in the background can lead to a best fit value for the number of $\eta$'s that is negative.
In this case we follow the suggestion of the Particle Data Group (PDG)~\cite{PDG:2006} and report a negative value with error bars for the cross section.
This provides sufficient information for constraints from these bins to be combined with nearby bins in comparing to theoretical predictions.

Acceptance for the $e p\to e \eta p$ reaction was calculated using 
a GEANT-based Monte Carlo simulation~\cite{GEANT}.
The event generator included radiative effects using the peaking
approximation~\cite{Ent01,MoTsai69}, and the cross sections have been 
corrected for radiation.  When making a major improvement in published
cross sections, development of an appropriate event generator is important.  
We use the data as a guide; the final cross section is dominated by $S$ and $P$
waves with the $S_{11}$(1535) the dominant structure seen.  An iterative 
procedure matching analyzed Monte Carlo to real data was used to
develop the event generator.  The same fitting procedure used on data 
was applied to Monte Carlo events to calculate acceptance.
The acceptance has significant variation across the bins with a maximum 
value of about $60\%$.
When approaching the kinematic limit, the acceptance falls off rapidly.
At the higher values of $W$, the proton goes forward where there is
a hole in the CLAS acceptance.   This causes problems for 
$\phi_\eta \sim 180^\circ$.
We only report results in bins where the acceptance is greater than $3\%$
and where it is not changing rapidly.

A detailed study of potential sources of systematic uncertainty was made.  Since
the $\eta$ peak shape and the background shape included various parameters,
all were studied.  The parameters were varied within the error bars determined 
in the fit for each cross section value.  Additional tests were made for 
variations in particle identification and in the momentum scale. 
The momentum uncertainty arose from uncertainty in the details of both
the magnetic field map and the alignments of the various tracking chambers.
Sensitivity to momentum determination was largest close
to threshold and falls off with increasing $W$.
Since most of the $\eta$ events are produced near threshold, this source dominates the average systematic uncertainty.
Cross sections
were recalculated with slightly tighter fiducial cuts and this variation
was considered as a systematic uncertainty estimate.
A variety of underlying physics models were used for evaluating the systematic on the radiative correction: using a single $S$-wave resonance or two, varying the mass and width of a single $S$-wave, or including a $P$-wave resonance.
The quoted systematic uncertainty on the radiative correction includes these 
effects, but is dominated by Monte Carlo statistics in the calculation.
The total systematic uncertainty for each bin in $W$, $Q^2$, and c.m. 
scattering angles was the sum of all the components added in quadrature.
The average total systematic uncertainty for the angle-integrated cross
sections was 3.3\%, 3.9\%, and 7.1\% for data at 1.5, 2.5 and 4.0
GeV, respectively.  The corresponding average
estimated systematic uncertainties for the differential cross sections
were 5.1\%, 5.2\%, and 7.6\%.  The breakdown by source for the 2.5 GeV data
(the set from which the largest number of data points come) is given 
in Table~\ref{tab:SysErrSum}.  
The estimated systematic uncertainties for individual data points were
seldom larger than the estimated statistical uncertainty. 

\begin{table}[!ht]
\begin{center}
\begin{tabular}{|l|c|c|} \hline
  Sys Err Source     & Angle-integrated (int)       & Differential (diff)     \\ \hline
 $M_{\eta}$          & $0.03\%$  & $0.06\%$ \\
 $\sigma_{\eta}$     & $0.4\%$   & $0.7\%$  \\
 $f$                 & $1.3\%$   & $1.4\%$  \\
 $\alpha$            & $0.1\%$   & $0.1\%$  \\
 $D_{bkg}(\msubx)$   & $0.1\%$   & $0.1\%$  \\ 
 fiducial cut        & $0.6\%$   & $2.3\%$  \\
 radiative corr.     & $1.0\%$   & $1.0\%$  \\
 momentum scale      & $2.6\%$   & $4.4\%$  \\ \hline
 total               & $3.9\%$   & $5.2\%$  \\ \hline
\end{tabular}
\caption{Summary of systematic uncertainties for the angle-integrated and differential cross section analysis.  $M_\eta$, $\sigma_\eta$, $f$, $\alpha$,
and ``radiative corr.'' describe the $\eta$ missing mass peak shape; $D_{bkg}$ is the background missing mass function; other entries parameterize various
detector properties.  See text for details.}
\label{tab:SysErrSum}
\end{center}
\end{table}


\section{\label{sec:rslt}Results}

\subsection{\label{sec:integ} Angle-integrated Cross Sections}

\begin{table}
\caption{\label{tab:tabint} Binning details for the angle-integrated cross sections.
For each $Q^2$ bin, we show the minimum and maximum values of $Q^2$, the energy of the electron beam for the data set, and the maximum value of $W$ probed.  The $W$ bin width in all cases was 10 MeV.
}
\begin{tabular}{|c|c|c|c|} \hline
\mbox{$Q^2_{min} (GeV^2)$} & $Q^2_{max} (GeV^2)$ & $E_{beam} (GeV)$ & $W_{max} (GeV)$\\
\hline
0.13 & 0.2 & 1.5 &1.66\\
0.2  & 0.3 & 1.5 &1.64\\
0.3  & 0.4 & 1.5 &1.61\\
0.6  & 0.8 & 2.5 &2.00\\
0.8  & 1.0 & 2.5 &1.90\\
1.0  & 1.2 & 2.5 &1.81\\
1.2  & 1.4 & 2.5 &1.69\\
1.3  & 1.7 & 4.0 &2.30\\
1.7  & 2.1 & 4.0 &2.30\\
2.1  & 2.5 & 4.0 &2.13\\
2.5  & 2.9 & 4.0 &1.93\\
2.9  & 3.3 & 4.0 &1.72\\ \hline
\end{tabular}
\end{table}
                                                                               
\begin{figure}[ht]
  \includegraphics[width=11cm]{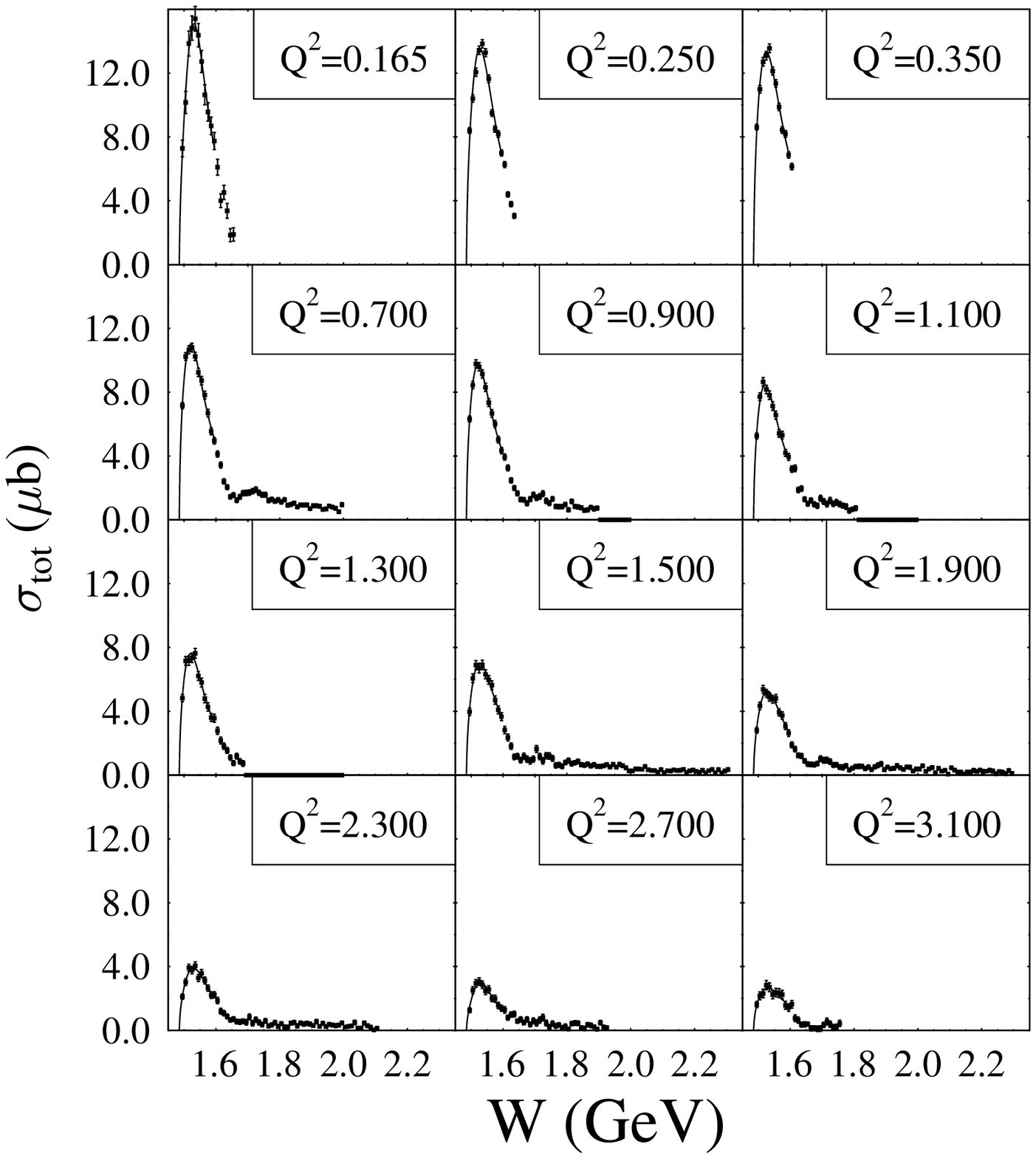}
  \caption{Angle-integrated cross sections ($\sigma(\gamma_v p \to \eta p$)) 
measured in all $Q^2$ bins. Only statistical uncertainties are included in the 
data points. Systematic uncertainties are small compared to statistical
errors and are not shown.  The line represents the single Breit-Wigner fit.}
  \label{fig:intxs}
\end{figure}
To get the angle-integrated cross sections, the events were binned in $W$ 
and $Q^2$, as shown in Table \ref{tab:tabint}.
The 1.5 GeV beam energy data covers the $Q^2$ range from 0.13 to 0.4 
GeV$^2$, while the upper two beam energies cover from 0.6 to 3.3 GeV$^2$.
Each bin is labeled by its centroid.  Results are tabulated in
the CLAS database~\cite{clasweb}.
These cross sections are presented in Fig.~\ref{fig:intxs}.
The prominent peak at $W\sim 1.5$ GeV is primarily populated through
intermediate excitation of the $S_{11}(1535)$ resonance.
Fits to a Breit-Wigner relativistic form with an energy-dependent 
width, Eq.~(\ref{eq:bw}) are used to fit the low $W$ region.
Various model calculations~\cite{Benmerrouche:1995uc} in the past 
have found a small 
nonresonant contribution to the cross section and none is needed
here.  The simple shape describes the low $W$ region well, but there 
are deviations for $W>1.6$ GeV, presumably due to interference
between $S_{11}$(1535), $S_{11}$(1650), and nonresonant processes.
Although the higher mass resonance is very near to the state
we seek to describe, all analyses~\cite{PDG:2006} find a
very small $\eta N$ branching fraction for $S_{11}$(1650). 
Therefore, we restrict the fit to $W$ values less than 1.6 GeV.
\begin{figure}[ht]
  \includegraphics[width=11cm]{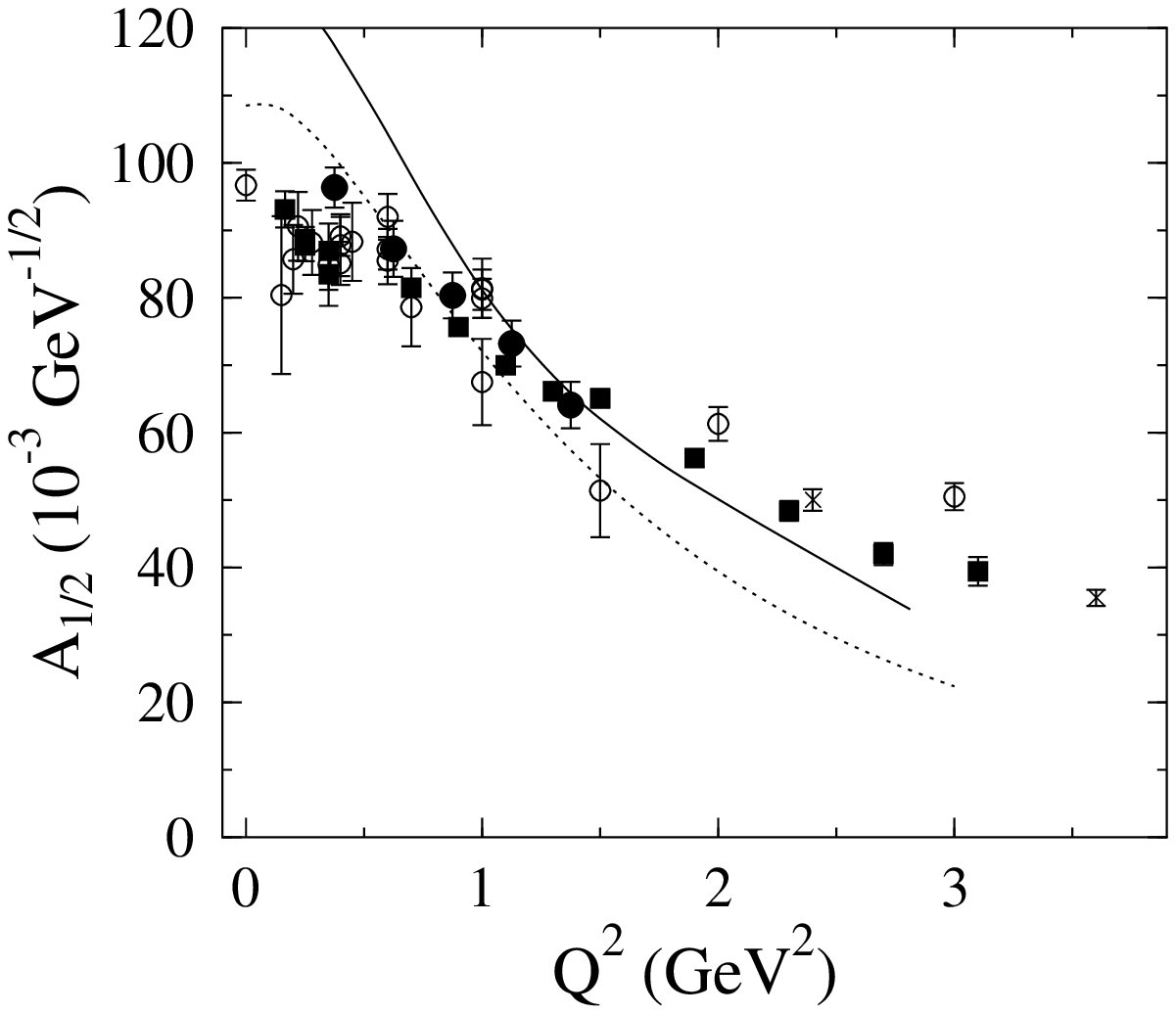}
  \caption{Extracted values for $A_{\frac{1}{2}}$ for 
$\gamma p \rightarrow\ S_{11}(1535)$ transition for this experiment 
(filled squares).  There are overlapping data points at $Q^2 = 0.25$ and 
$0.35$ GeV$^2$ coming from data taken with two different magnetic field 
settings. Filled circles show results from our previous smaller 
data set~\cite{Thompson:2000by}.  JLab results from 
Armstrong \etal~\cite{Armstrong:1998wg} are shown as crosses. 
The open circles are results from earlier publications~\cite{Kummer:1973em,Beck:1974wd,Brasse:1978as,Breuker:1978qr,Brasse:1984vm,Krusche:1995nv}. 
All results have been converted to represent a common width and 
branching ratio for $S_{11}(1535) \rightarrow \eta p$. The theoretical 
models are from constituent 
quark models of Capstick and Keister~\cite{Capstick:1995ne} (solid line) 
and Aiello, Gianinni, and Santopinto~\cite{Aiello:1998xq} (dotted curve).}
  \label{fig:a12}
\end{figure}
Two previous experiments~\cite{Brasse:1978as,Breuker:1978qr} performed 
longitudinal/transverse separations in the late 1970's.
Their results are consistent with no longitudinal component, albeit 
with large uncertainties.  For the results presented here, the different 
beam energies have insufficient overlap in $W$ and $Q^2$ to separate 
these components.  Under the assumption that the cross section is 
dominated by a single resonance and that $S_{\frac{1}{2}}$ is small, 
we can relate the $A_{\frac{1}{2}}$ to the peak cross sections extracted 
from the fit (see Eqs.~(\ref{eq:xs}),(\ref{eq:a12})):
\begin{equation}
  A_{\frac{1}{2}}(Q^2)=\left[\frac{W_R\Gamma_R}{2 m_p b_\eta} \sigma(W_R,Q^2)\right]^{1/2}.
\end{equation}
Consistent with Armstrong \etal~\cite{Armstrong:1998wg}, a value of
 the full width of 150 MeV and an $S_{11}\to\eta N$ branching 
 ratio of 0.55 were used. 
The results of this determination of $A_{\frac{1}{2}}$ are shown in 
Fig.~\ref{fig:a12} along with some previous results converted to be 
consistent with our choice of $\Gamma_R$ and $b_\eta$.
The precise normalization of $A_{\frac{1}{2}}$ depends on the choice 
of parameters for the contributing resonances, which are, as yet, 
not well determined.  For instance, using the range of values listed 
in PDG for $\Gamma_R$ and $b_\eta$ leads to a $11\%$ systematic uncertainty 
on $A_{\frac{1}{2}}$.  While these uncertainties affect the {\em absolute} 
value of $A_{\frac{1}{2}}$, the {\em shape} of the $Q^2$ dependence is 
much better determined.
More detailed understanding of this state is required to better 
determine absolute values of $A_{\frac{1}{2}}$ and estimate the model 
dependence of those values.  Using the choice of $\Gamma$ and $b_\eta$ 
described above, our extracted values are 
consistent with the previous values at low $Q^2$~\cite{Kummer:1973em,Beck:1974wd,Brasse:1978as,Breuker:1978qr}, but with smaller 
uncertainties.  At high $Q^2$, there is moderate disagreement between 
the previously published results of Brasse \etal~\cite{Brasse:1984vm} 
(at $Q^2=2.0$ and 3.0 GeV$^2$) and Armstrong \etal~\cite{Armstrong:1998wg} 
(at 2.4 and 3.6 GeV$^2$).  Our results match up nicely with 
Armstrong \etal\ and provide a precise determination of the shape of 
the $Q^2$ dependence of $A_{\frac{1}{2}}$ from low $Q^2$ up to their 
high $Q^2$ determinations.

The literature has various theoretical calculations of the photon
coupling amplitude within the Constituent Quark Model (CQM).  Matching 
the slow falloff with $Q^2$ has been difficult.  We show two
recent calculations~\cite{Capstick:1995ne,Aiello:1998xq}.
Aiello, Giannini, and Santopinto~\cite{Aiello:1998xq} use a
hypercentral CQM and emphasize the importance of the
3-body quark force.  Although this prediction gives the
 best agreement with our data of all the calculations, it falls 
off more rapidly with  $Q^2$ than the data.  The Capstick and
Keister calculation~\cite{Capstick:1995ne} starts with the more
traditional CQM, but use relativistic dynamics in a light-front
framework.  Although the two calculations use
different approaches, the CQM is not well-defined and many other
results are given in the literature.

\subsection{\label{sec:differ}Differential Cross Sections}

\begin{table}
\caption{\label{tab:tabdif}
Binning details for the differential cross sections.
For each $Q^2$ bin, we show the minimum and maximum values of $Q^2$, 
the energy of the electron beam for the data set, the maximum value of 
$W$ probed, and the bin width in $\cos{\theta^*}$.  
The $W$ bin widths in all cases were 20 MeV, while the $\phi*$ bins were 
$30^\circ$ wide.
}
\begin{ruledtabular}
\begin{tabular}{ccccc}
$Q^2_{min}$ & $Q^2_{max}$ & $E_{beam}$ & $W_{max}$ & $\Delta\cos{\theta^*}$\\
(GeV$^2$)   & (GeV$^2$)   & (GeV)      &  (GeV)    & \\
\hline
0.2  & 0.4 & 1.5 &1.60 & 0.2\\
0.6  & 1.0 & 2.5 &1.80 & 0.2\\
1.0  & 1.4 & 2.5 &1.74 & 0.2\\
1.3  & 2.1 & 4.0 &2.00 & 0.4\\
2.1  & 2.9 & 4.0 &1.92 & 0.4\\
\end{tabular}
\end{ruledtabular}
\end{table}
                                                                               
\begin{figure}
 \includegraphics[width=12cm,angle=-90]{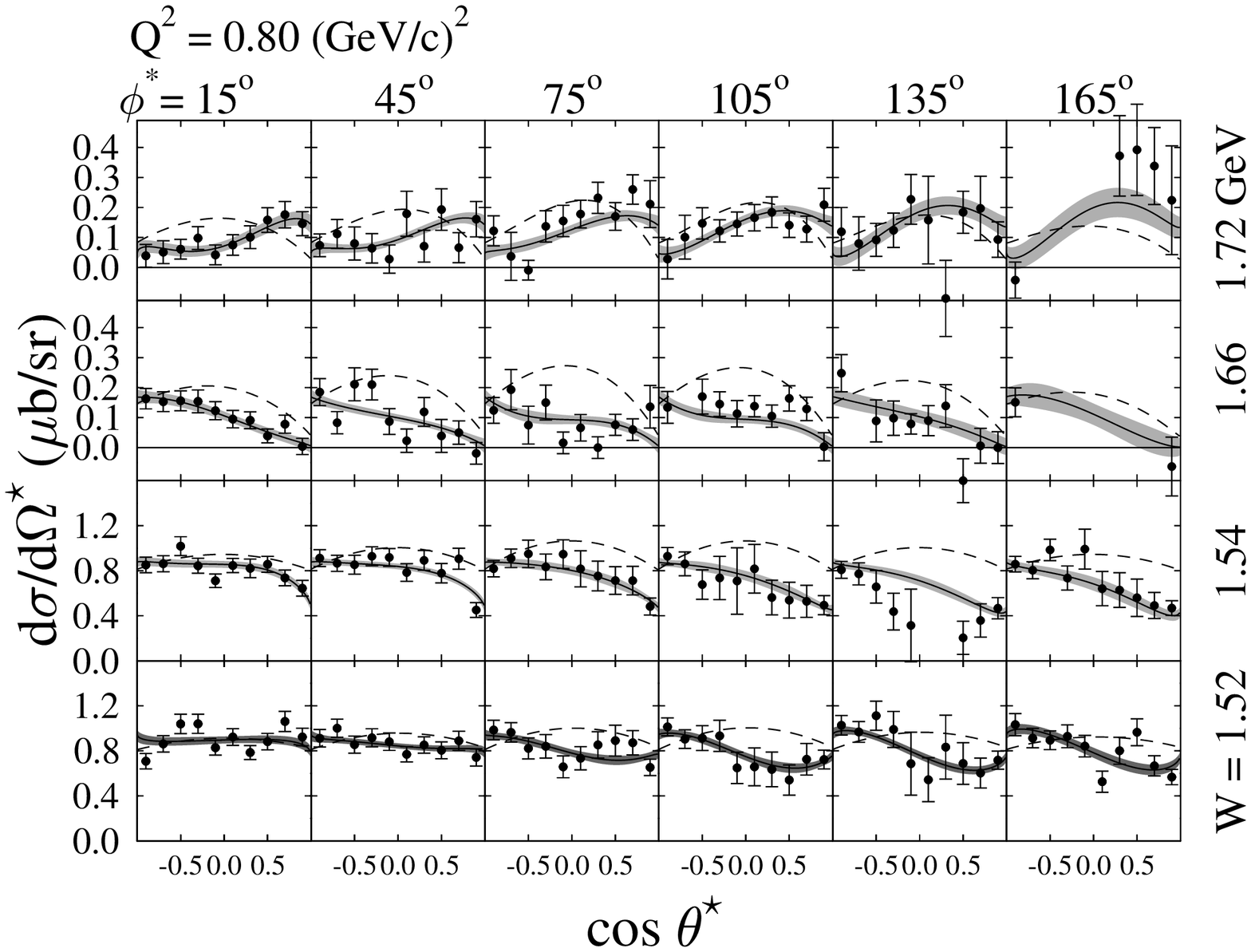}
  \caption{Sample differential cross sections for $\gamma_v p \to \eta p$
in the center of mass frame for
$Q^2= 0.8$ GeV$^2$ and selected $W$ bins.
Values for $\phieta$ symmetric about $180^\circ$ have been averaged.  (No
information is lost this way; see Eq.~(\ref{eqn:equ1}).)  Solid lines with 
an error band correspond to the response function fit described in the text.
Dashed  lines correspond to the calculation of $\eta$-MAID~\cite{Chiang:2001as}.}
  \label{fig:diffxs}
\end{figure}
For larger bins in $W$ and $Q^2$ (see Table \ref{tab:tabdif}), we extract
 differential cross sections versus center-of-mass scattering angles of the
 $\eta$ ($\cos\theta^*$ and $\phi^*$). Each bin is labeled by its centroid.
Results from this experiment are tabulated in the CLAS database~\cite{clasweb}.
For the $Q^2=0.8$ GeV$^2$ bin, Figure \ref{fig:diffxs} shows sample cross sections for
four $W$ bins.
The first two $W$ bins, $W=1.52$ and $1.54$  GeV are at the peak of the $S_{11}(1535)$ resonance.
They show a dominant isotropic component due to the $S_{11}(1535)\to\eta p$ process, but deviations from isotropy can be seen, especially at large $\phi^*$.
By $W=1.66$ GeV, the non-isotropy is quite evident.
The cross section falls monotonically as a function of $\cos\theta^*$, with the cross section for forward $\eta$ production consistent with zero.
As $W$ increases this feature changes dramatically.
At $W=1.72$ GeV, the forward-backward asymmetry of the distributions has reversed, with forward $\eta$ production favored, while backward production is close to zero.
  
The $\eta$-MAID model~\cite{Chiang:2001as}, based on the MAID 
formalism~\cite{Drechsel:1998hk}, has been developed for $\eta$ electro- 
and photoproduction.  This is an isobar model using a relativistic
Breit-Wigner $W$ dependence with form factors.  Eight PDG 3* and 4* resonances
of mass less than 1.8 GeV and nonresonant processes are included at the
amplitude level.  They fit the photoproduction 
data~\cite{Krusche:1995nv,Ajaka:1998zi,Renard:2000iv} and the $Q^2$ 
dependence of the total cross section from electroproduction 
data~\cite{Thompson:2000by,Armstrong:1998wg} in the $S_{11}(1535)$ region.
The results of a calculation implementing this model are included in 
Figure \ref{fig:diffxs}.  These calculations roughly match the observed 
cross sections.  However, the angular dependence predicted by 
$\eta$-MAID does not agree with our data at $W$ above the $S_{11}(1535)$
region.  The model was not
fit to the differential cross sections of our previous work and
the $Q^2$ dependence of the higher mass resonances, e.g. $D_{15}$(1675),
was taken from a quark model calculation rather than from data.

For each $W$ and $Q^2$ bin, the differential cross sections are fit to a form
 that comes from an expansion of the response functions from
 Eq.~(\ref{eqn:equ1}) in terms of associated
 Legendre polynomials ($P_\ell^m(\coseta )$),
\begin{eqnarray}\label{equ2}
  \frac{d^{2}\sigma}{d\Omega_{\eta}^{*}}
   &=&  
    \sum_{\ell=0}^\infty {A_\ell} { P_\ell^0(\coseta )}
     +\nonumber\\ & & \sum_{\ell=1}^\infty{B_\ell} {P_\ell^1(\coseta )} \cos\phi_\eta^* 
       +\nonumber\\ & & \sum_{\ell=2}^\infty {C_\ell} {P_\ell^2(\coseta )} \cos 2\phi_\eta^*.
\end{eqnarray}
The parameters $B_\ell$, $B_\ell$, and $C_\ell$ depend on $Q^2$, $W$, and
$\epsilon$.  They represent
bilinear sums over contributing multipole amplitudes. 
Truncating to $\ell\le 3$, we determined the parameters in 
Eq.~(\ref{equ2}) by a fit to the data.
This truncation is motivated by three effects.
i) The lightest known $N^*$ resonance with $\ell>3$ is the $G_{17}(2190)$;
if the dominant effects on the differential cross section arise from 
interference with the dominant $\ell=0$ partial wave, terms above $l=3$ 
should be negligible.  ii) Fits to the $\eta$-MAID predicted cross sections 
yield negligible contributions for terms higher than $\ell=3$.
iii) Good fits to the data are obtained with the truncated sum.
\begin{figure}[ht]
  \includegraphics[width=14.0cm]{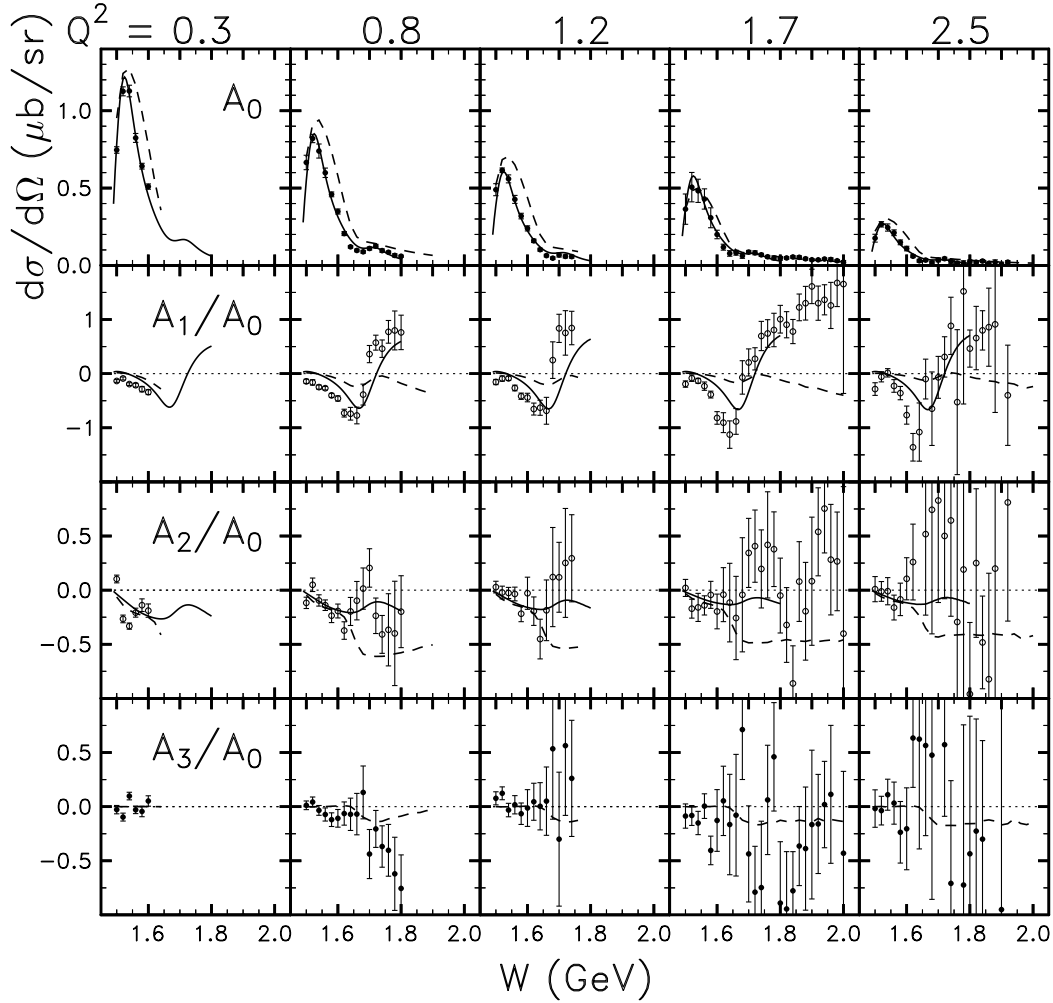}
  \caption{Results from fitting the angular distribution data of this 
experiment to Eq.~(\ref{equ2}). 
Coefficients of the $\phi^*$ independent terms are shown, i.e.
those that contribute to $\sigma_T+\epsilon\sigma_L$. 
Contributions from both statistical and systematic sources are displayed.
The dashed line is the $\eta$-MAID prediction~\cite{Chiang:2001as} and
the solid line is a four resonance fit to these terms.}
  \label{fig:parmsfit_a}
\end{figure}
\begin{figure}[ht]
  \includegraphics[width=14.0cm]{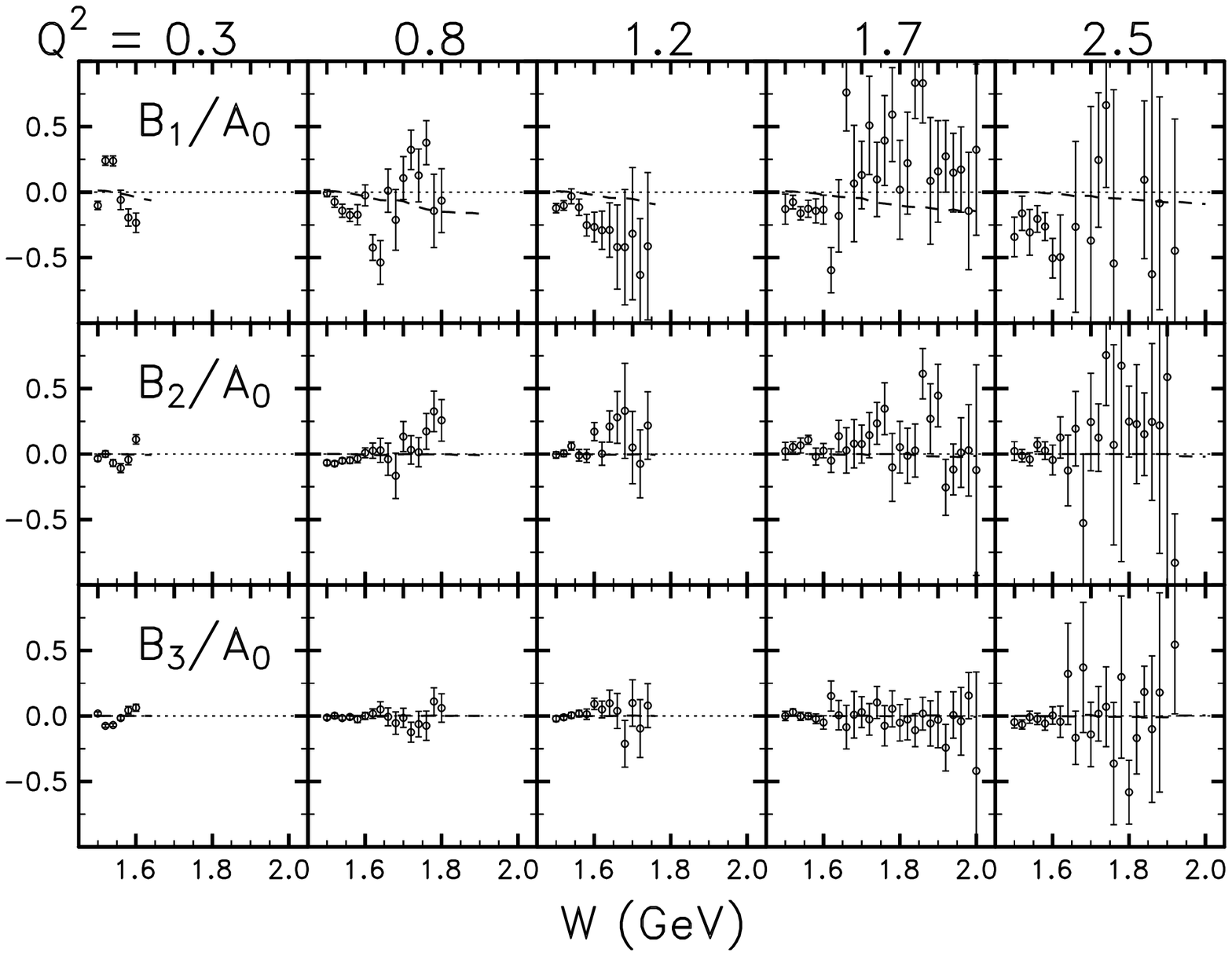}
  \caption{Same as Figure \ref{fig:parmsfit_a}, but showing the parameters corresponding to $\sigma_{LT}$.  For the four resonance fit, these parameters
are all zero.}
  \label{fig:parmsfit_b}
\end{figure}
\begin{figure}[ht]
  \includegraphics[width=14.0cm]{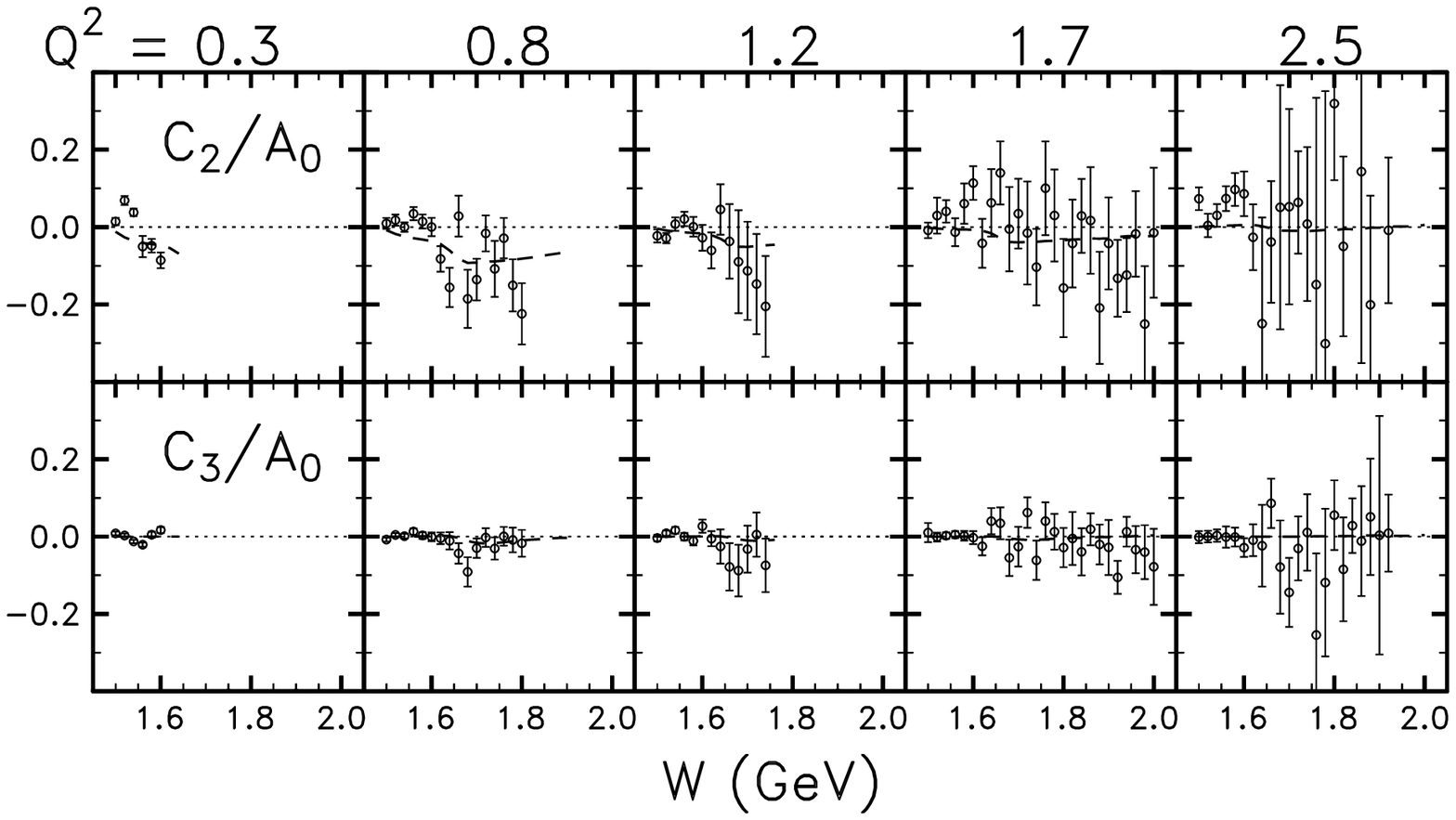}
  \caption{Same as Figure \ref{fig:parmsfit_a}, but showing the parameters corresponding to $\sigma_{TT}$.  For the four resonance fit, these parameters
are all zero.}
  \label{fig:parmsfit_c}
\end{figure}
  
Results are shown versus $W$ in Figures \ref{fig:parmsfit_a}, 
\ref{fig:parmsfit_b}, and \ref{fig:parmsfit_c}.
The quoted uncertainties contain both statistical and systematic uncertainties.
We repeated the fit taking into account shifts in the cross section 
for each of the sources of systematic uncertainty studied in Section \ref{sec:anal}. 
The total systematic uncertainty on the extracted parameters is the sum 
in quadrature of all individual sources.
We normalize our fitted $A_1$, $A_2$, etc. to the isotropic
term ($A_0$) in order to more clearly show the $W$ and $Q^2$ dependence 
of the shape of the differential cross section.  For the ratios the resulting 
uncertainty is dominated by the uncertainty on the numerators.

The isotropic component ($A_0=\sigma_{tot}/4\pi$) shows the same features 
as the angle-integrated cross sections: a dominant peak from the $S_{11}(1535)$ 
with additional structure above $W=1.6$ GeV.
The other prominent term in the fit is $A_1$, which represents the slope of
 the differential cross section versus $\coseta$.
A structure in the $W$ dependence of $A_1$  was first seen in our 
previous publication~\cite{Thompson:2000by}
and is also  seen by the GRAAL photoproduction 
experiment~\cite{Renard:2000iv}. 
By examining the ratio $A_1/A_0$ in the new data, we can study this 
structure in more detail.  Two features stand out in this ratio:
\begin{enumerate}
\item The ratio $A_1/A_0$ is large and makes a rapid change from negative to positive values at $W\approx 1.7$ GeV.
\item This structure is roughly independent of $Q^2$ up to 2.5 GeV$^2$.
\end{enumerate}
The simplest description for $A_1$ is in terms of interference between $S$ and
 $P$ waves.
In that case the rapid change in $A_1$ between $W=1.66$ and $1.72$ GeV could be
 caused by one of the waves passing through a resonance.  
There are 2 $P$-wave resonances in this region, $P_{11}(1710)$ and $P_{13}(1720)$.
The former is rated $3*$ by the PDG~\cite{PDG:2006}, but its properties are
very difficult to extract from data and is therefore 
controversial~\cite{Arndt:2003ga}.  The latter is rated $4*$, but is
also poorly understood~\cite{Vrana:1999nt}.  Fits to CLAS $\pi^+ \pi^-$ 
electroproduction data~\cite{Ripani:2002ss} provided evidence
that the existing baryon structure at $W \sim$ 1.7 GeV should be changed.  
Their fits prefer either a greatly reduced $\rho N$ decay branch for
the existing $P_{13}$(1720) resonance or a new $\frac{3}{2}^+$ state.
In the present data, we cannot couple to a $T=3/2$ state and
are unable to distinguish between $P_{11}$ and $P_{13}$ states; we 
choose to use only a $P_{11}$
state.  If one describes the cross section using {\em only} $S_{11}$ 
and $P_{11}$ partial waves,
\begin{equation}
  \frac{A_1}{A_0}= \frac{2\Re(E_{0+}^*M_{1-})}{|E_{0+}|^2+|M_{1-}|^2}.
\end{equation}
In this case, the rapid shift from backward to forward peaked cross 
sections would be due to a rapid change in the relative phase of the 
$E_{0+}$ and $M_{1-}$ multipoles because one of them is passing through 
resonance.  The observation that this structure in $A_1/A_0$ is 
approximately $Q^2$ independent would then imply that $S_{11}$ 
and $P_{11}$ partial waves have a similar $Q^2$ dependence.

The values of $B_\ell$ shown in Fig.~\ref{fig:parmsfit_b} are 
consistent with zero.  These parameters measure the $\sigma_{LT}$ component 
of $\frac{d\sigma}{d\Omega}$, indicating that longitudinal amplitudes 
are not significant for this reaction (as was assumed in 
Section~\ref{sec:integ}).  The $C_l$ parameters in Fig.~\ref{fig:parmsfit_c} 
measure the $\sigma_{TT}$ component of $\frac{d\sigma}{d\Omega}$.
They are small indicating the $A_{3/2}$ components are also small 
for these values of $Q^2$.

To better understand the content of the $\eta$-MAID model, we also fit 
the parameters in Eq.~(\ref{equ2}) to the predicted cross sections from 
that model.  The extracted parameters are also included in
 Figures \ref{fig:parmsfit_a}, \ref{fig:parmsfit_b}, and \ref{fig:parmsfit_c}
as dashed lines.
The prediction has a broader $S_{11}$ peak than is seen in our data.
Some structure is predicted in $A_1$ arising from the $P_{11}(1710)$,
but the size of this effect is not nearly enough to match our data.
At high $W$, $\eta$-MAID predicts a negative $A_1$ in contrast to the 
significant positive value we observe.  The model value 
of $b_{\eta N}$ for $P_{11}$(1710) is much larger than the PDG value.
Our data indicates the model value is incorrect.
$\eta$-MAID contains many sources of $D$-wave contributions: $D_{13}(1520)$, 
$D_{13}(1700)$, and $D_{15}(1675)$ in addition to nonresonant amplitudes.
This produces a value for $A_2/A_0$ that matches our data for $W<1.6$ GeV, 
where the $D_{13}(1520)$ is the leading contribution.  At larger $W$,
agreement is poor.  The model value of $b_{\eta N}$ for $D_{15}$(1675) 
is much larger than the PDG value; our data indicates this is incorrect.
The prediction for $A_3/A_0$ is near zero, as are our measurements.

Predictions for the $\sigma_{LT}$ and $\sigma_{TT}$ terms are consistent with our measurements.
$C_2$ is the only term that is not negligible in $\eta$-MAID for our values of $Q^2$.
It arises from the $A_{3/2}$ amplitudes of the $D$-wave resonances interfering with the larger $S$-wave amplitude.   
Our data agrees with this general trend, but the effect is small compared to the uncertainties.

To gain further understanding of the resonance content of our data,
we did an additional fit to the differential cross section data 
using relativistic Breit-Wigner resonances
according to Eqs.~(\ref{eq:sigbw})-(\ref{eq:width}).
We fit the extracted parameters, up to $W=1.8$ GeV, to a sum of four amplitudes for the following resonances: $S_{11}(1525)$, $S_{11}(1650)$, $P_{11}(1710)$, and $D_{13}(1520)$.  This set of resonances was determined empirically
as the minimal set required to fit the general features of our data.  
Although the
properties of $P_{11}(1710)$ are very uncertain, it is an important
contributor to this fit.  We label it as $P_{11}$, but we cannot
distinguish between $P_{11}$ and $P_{13}$(1720) in our data set;
specifically, a $P_{13}$ resonance would also give a rapid energy
dependence in either $A_2$ or $C_2$ which we are unable to exclude
with current statistical accuracy.  Only the transverse response function,
$\sigma_T$ is modeled, i.e. the $A_i$ parameters.  For the resonances with 
small contributions ($S_{11}(1650)$ and $D_{13}(1520)$), we fixed the 
resonance parameters to values 
obtained elsewhere.  Masses and widths were set to average values 
from the Particle Data Group.  For the $D_{13}(1520)$ we used the $Q^2$ 
dependence of $\eta$-MAID.  Following the assumption of the single quark 
transition model~\cite{Hey:1974}, the ratio of the strength of the $S_{11}(1650)$ to that of the $S_{11}(1535)$ was taken to be independent of $Q^2$.
Motivated by the $Q^2$ independence of $A_1/A_0$ in our data, we 
assumed the $P_{11}(1710)$ had the same $Q^2$ dependence as the 
$S_{11}$ states.  This left 12 variables in the fit:  the masses and 
widths of the $S_{11}(1535)$ and $P_{11}(1710)$, the relative strengths 
of the $S_{11}(1650)$ and $P_{11}(1710)$ to that of the $S_{11}(1535)$, 
an overall strength of the $D_{13}(1520)$, and the absolute strength 
of the $S_{11}(1535)$ in each of the five $Q^2$ bins.
We view this as a simple fit.  Our results should not be interpreted as 
a precise determination of resonance parameters, but rather as an 
indication of the dominant components needed in any future theoretical work.  

The results of this fit are also shown in Figure \ref{fig:parmsfit_a}.
The fit yields a reasonable, though not perfect description of our data.
The isotropic term $A_0$ is described by the dominant $S_{11}(1535)$ peak, 
modified by the smaller $S_{11}(1650)$.
The deviation from a simple Breit-Wigner is described as a combination 
of destructive interference between the $S_{11}(1535)$ and $S_{11}(1650)$, 
and a small contribution from the $P_{11}(1710)$.
Including the $S_{11}(1650)$ results in an extracted value of 
$A_{\frac{1}{2}}$ for the $S_{11}(1535)$ which is $7\%$ higher than that 
obtained with a single Breit-Wigner.  The fitted width of the $P_{11}(1710)$ 
is 100 MeV, which is consistent with the central (but very uncertain) 
PDG value.
We cannot isolate the $P_{11}(1710)$ photocoupling from that 
state's branching ratio into $\eta p$; we can only quote a ratio of 
$\xi$ values (Eq.~(\ref{equ:xi})).
The extracted value of $\xi_{1710}/\xi_{1535}$ is 0.22, which is about twice as large as in $\eta$-MAID, and nearly an order of magnitude larger than that extracted from parameters of the $P_{11}(1710)$ in the PDG.
The $D_{13}(1520)$ primarily effects the quadratic term $A_2$.  Including this resonance is enough to give a reasonable description of the $W$ dependence of $A_2$.
Our data do not require significant contributions from higher $D$-wave states  present in the $\eta$-MAID model.

We fit the structure in $A_1/A_0$ with a smooth $S$-wave and a rapidly 
changing $P$-wave.  One could also describe this structure in terms of a 
new $S$-wave resonance interfering with $P$-wave component as in the model 
of Saghai and Li~\cite{Saghai:2001yd}.  However, the amplitudes for the new 
resonance and the $P$-wave component must both fall off slowly with $Q^2$ 
to reproduce the data.
 
We also fit the $\phi$ dependence of the differential cross sections directly 
to Eq.~(\ref{eqn:equ1}) in order to obtain $\sigma_T + \epsilon\sigma_L$, 
$\sigma_{LT}$, and $\sigma_{TT}$ as a function of $W$, $Q^2$ and 
$\cos{\theta^*}$.  We choose to fit the $\phi^*$ dependence in terms of 
the parallel/perpendicular asymmetry (Asym$_{TT}$) and the 
parallel/anti-parallel asymmetry (Asym$_{LT}$).
\begin{equation}
\mathrm{Asym}_{TT} = \frac{\sigma_{||} - \sigma_\perp}{\sigma_{||} + \sigma_\perp},
\end{equation}
where
\begin{equation}
\sigma_{||} = \frac{1}{2}\left (\sigma(\phi=0)+\sigma(\phi=\pi)\right )
\end{equation}
and
\begin{equation}
\sigma_\perp = \frac{1}{2}\left (\sigma(\phi=\pi/2)+\sigma(\phi=3\pi/2)\right
).
\end{equation}
\begin{equation}
\mathrm{Asym}_{LT} = \frac{\sigma(\phi=0) - \sigma(\phi=\pi)}{\sigma(\phi=0) +
\sigma(\phi=\pi)}.
\end{equation}

\begin{figure}[ht]
  \includegraphics[width=12.0cm]{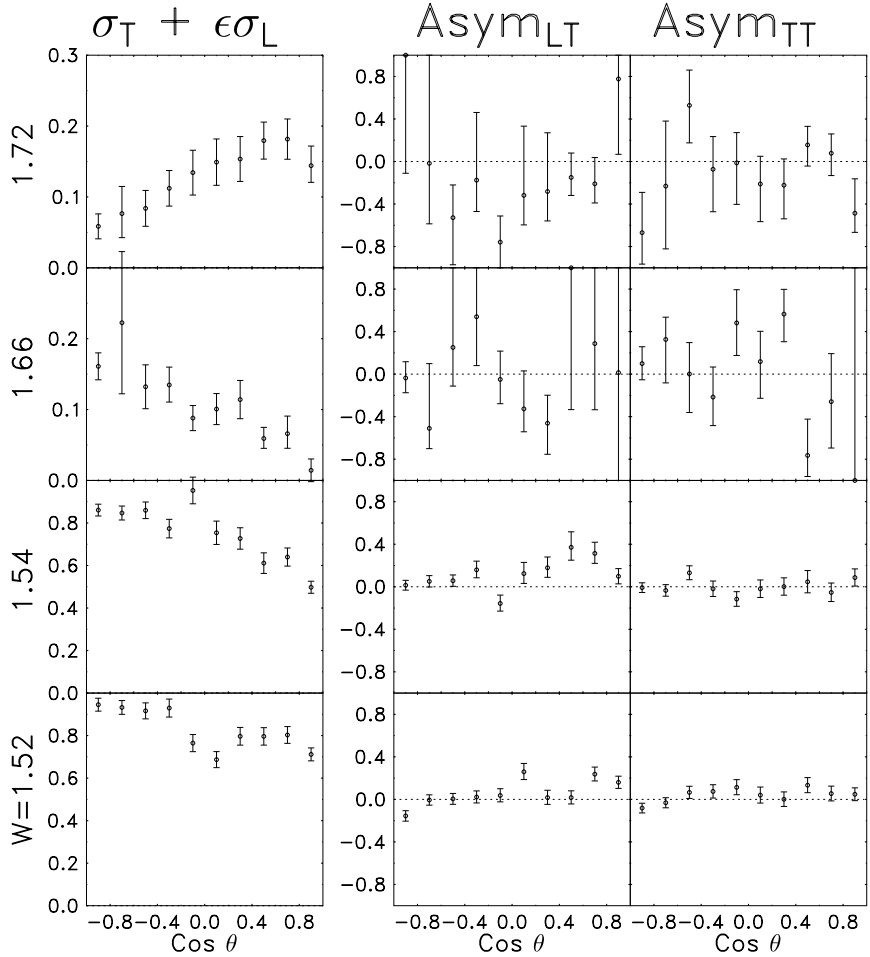}
  \caption{Extracted values for $\sigma_T + \epsilon\sigma_L$, Asym$_{LT}$, and Asym$_{LT}$ as a function of $\cos{\theta^*}$ for four selected $W$ bins with $Q^2=0.8$ GeV$^2$. }
  \label{fig:response}
\end{figure}
For photoproduction, $\sigma_L$ and $\sigma_{LT}$ don't contribute.  A common
polarization parameter is the parallel/perpendicular asymmetry, $\Sigma$.
It is defined by $\mathrm{Asym}_{TT} = \epsilon\Sigma$; note also that
$\Sigma = \sigma_{TT}/\sigma_T$. In electroproduction, the possible 
presence of a longitudinal term makes the relationships more complicated:
\begin{equation}
\mathrm{Asym}_{TT} = \frac{\epsilon\sigma_{TT}}{\sigma_T + \epsilon\sigma_L}
\end{equation}
\begin{equation}
\mathrm{Asym}_{LT} = \frac{\sqrt{2\epsilon(\epsilon+1)}\sigma_{LT}}
{\sigma_T+\epsilon\sigma_L+\epsilon\sigma_{TT}}.
\end{equation}
The data was analyzed in terms of these 3 response function combinations.
A treatment of systematic uncertainties similar to that used for differential
cross sections was applied. 
Figure \ref{fig:response} shows the values extracted from these fits for the 
same $W$ and $Q^2$ shown in Figure \ref{fig:diffxs}.  Error bars
display the systematic and statistical uncertainties.
The quantity $\sigma_T + \epsilon\sigma_L$, shows the same features we have 
discussed earlier.  The values for Asym$_{TT}$ are consistent with 
zero in all distributions, but the size of the estimated error bars are a strong
function of $W$.  For $W>1.6$, the total cross section is smaller than
where the $S_{11}$ resonance dominates.   Extraction of  meaningful values 
for the $\phi$ dependence in this manner is therefore difficult.

\section{\label{sec:disc}Conclusions}
Our extractions of $A_{\frac{1}{2}}$ for the excitation of the $S_{11}(1535)$ 
cover a large range and match up well with Armstrong's 
results~\cite{Armstrong:1998wg} at higher $Q^2$.
It should be noted again that there are significant model dependencies on describing the mass, width and branching ratio into $\eta p$.
These uncertainties lead to significant systematic uncertainties on the absolute scale of $A_{\frac{1}{2}}$.
These uncertainties are common to all points currently determined, so the shape of the distribution is well determined.
It becomes a significant challenge for theory to reproduce this shape.
No existing model is able to describe the full range.


Knowledge of the $N^*$ resonances in the region $W \sim$ 1700 MeV is 
presently weak because the quality of older $\pi N \rightarrow \pi \pi N$ 
and $\pi N \rightarrow \eta N$ data is poor.  The coupling of known 
$P$-wave resonances to $\eta N$ is thought to be very small.  In this 
experiment, rapid energy dependence in the strength in the $P$-wave for
coupling to $\eta N$ final states is found.  With a simple resonance
model, we are able to describe these data with 
significant coupling of a $P$-wave resonance to $\eta N$.  As
with $S_{11}$(1535), the falloff of this coupling must be
very slow.

Although we can describe our measurements in terms of the $P_{11}(1710)$, we 
cannot distinguish between that or the $P_{13}(1720)$ with these data .
Either resonance could produce the effect seen in $A_1/A_0$. 
The $P_{11}(1710)$ is more poorly understood than the $P_{13}(1720)$, so 
it is easier to accommodate our data by altering the partial widths of 
the $P_{11}$ rather than the $P_{13}$.
A large $P_{13}$ could also produce effects in other terms.
For instance, interference with a $D$-wave, would give a small 
contribution to $A_{\frac{3}{2}}$, but
not significant compared to our uncertainties.
A $P_{13}$ resonance would have an $A_{\frac{3}{2}}$ photo-excitation as well as $A_{\frac{1}{2}}$.
Our determination of $A_1/A_0$ is sensitive to the $A_{\frac{1}{2}}$ amplitude;
 a significant $A_{\frac{3}{2}}$ amplitude could also lead to large effects in $\sigma_{TT}$.

Evidence for possible alterations in $N^*$ $P$-wave resonances at
masses about 1.7 GeV is accumulating.  In addition to what is found
in this experiment, double-pion production experiments in this same mass 
range~\cite{Ripani:2002ss,Assafiri:2003mv} are also unable to
be described with models using existing information.  Since different models
are used to describe the different data sets, it is important to
use a common model to describe the combined measurements from these 
(and other) reactions.  Such a program may allow us to accurately 
determine the properties of the $P$-wave resonances in this region.

\section{\label{sec:ack}Acknowledgments}
We acknowledge the outstanding efforts of the staff of 
the Accelerator and the Physics Divisions at JLab that made this experiment 
possible. This work was supported in part by the U.S. Department of Energy, 
the National Science Foundation, the Istituto Nationale di Fisica Nucleare,
the French Centre National de la Recherche Scientifique, the French 
Commissariat \`a l'Energie Atomique, 
and the Korea Science and Engineering Foundation.  Jefferson Science
Associates (JSA) operates the Thomas Jefferson Science
Facility for the United States Department of Energy under contract
DE-AC05-06OR23177.

\bibliography{eta-refs}

\end{document}